\documentclass[12]{article}
\usepackage{fleqn}
\usepackage{graphicx}
\begin{document}
\Large
\begin{center}
{\bf On Cremonian Dimensions  Qualitatively Different from Time
and Space}
\end{center}
\vspace*{-.3cm} \Large
\begin{center}
Metod Saniga
\end{center}
\vspace*{-.5cm} \small
\begin{center}
{\it Astronomical Institute, Slovak Academy of Sciences, 05960
Tatransk\' a Lomnica, Slovak Republic}

\vspace*{.0cm}
 and

\vspace*{.0cm} {\it Institut FEMTO-ST, CNRS, Laboratoire de
Physique et M\'etrologie des Oscillateurs,\\ 32 Avenue de
l'Observatoire, F-25044 Besan\c con, France}
\end{center}

\vspace*{-.3cm} \noindent \hrulefill

\vspace*{.0cm} \small \noindent {\bf Abstract}

\noindent We examine a particular kind of six-dimensional
Cremonian universe featuring one dimension of space, three
dimensions of time and other two dimensions that can{\it not} be
ranked as either time or space. One of these two, generated by a
one-parametric aggregate of (straight-)lines lying on a quadratic
cone, is more similar to the spatial dimension. The other,
represented by a singly-parametrical set of singular space quartic
curves situated on a proper ruled quadric surface, bears more
resemblance to time. Yet, the two dimensions differ profoundly
from both time and space because, although being macroscopic, they
are {\it not} accessible to (detectable by) {\it every} Cremonian
observer. This toy-model thus demonstrates that there might exist
extra-dimensions that need not necessarily be compactified to
remain unobservable.

\vspace*{-.1cm} \noindent \hrulefill

\vspace*{.3cm} \normalsize \noindent There are a number of
features of the macroscopic physical world that still remain
substantially beyond grasp of theoretical physics. Among them, the
non-trivial structure of time and the observed dimensionality of
the universe obviously represent a case in question. As we  found
[1,2] and have repeatedly stressed [3--5], the two properties seem
to be intimately intertwined and ask, therefore, for a
conceptually new approach to be properly understood. A (very
promising) piece of such a formalism is undoubtedly the
concept/theory of Cremonian space-times [6--14].

The Cremonian picture of space-time is indeed remarkable in
several aspects. The first, and perhaps most notable, fact is that
without employing any concept of metric (measure), this approach
fundamentally distinguishes the time dimension(s) from spatial
ones and, in its most trivial form, it straightforwardly leads to
their observed number (4) and respective ratio (1+3) as well
[6,7,9,10]. Second, it demonstrates that these dimensions are not
primordial, but emerge from more fundamental algebraic geometrical
structures [13]. Third, it indicates that the universe with the
inverse signature might evolutionary be intimately connected with
our universe [12]. And last, but not least, when the observer
(subject) is concerned, it qualitatively reproduces our ordinary
perception of time as well as a whole variety of
altered/non-ordinary forms of mental space-times [10,15];
moreover, every observer in this basic Cremonian universe is found
to face an intricate 2+1 break-up among the space dimensions
themselves [14].

In this paper, we introduce and examine a particular kind of a
more complex, six dimensional Cremonian universe whose
spatio-temporal sector is still four dimensional, yet featuring
three dimensions of time and just a single spatial coordinate. The
character of other two dimensions is neither that of space nor
time; in addition, these dimensions are only conditionally
observable/accessible. This Cremonian universe sits in the
3-dimensional projective space over the fields of the real numbers
$\Re$ and is generated by the configuration of fundamental
elements of a homaloidal web of cubic (i.e., third-order) surfaces
that share a proper conic, ${\cal \widehat{Q}}$, a
(straight-)line, ${\cal \widehat{L}}$, incident with the conic and
not lying in its plane, and three different non-collinear points,
$\widehat{B}_i$ ($i$=1,2,3), none of them incident with either
${\cal \widehat{L}}$ or the plane of ${\cal \widehat{Q}}$. The
cubics of the web have also two double points, $D_1$ and $D_2$, in
common; both the points lie on ${\cal \widehat{L}}$, the former
being the intersection of ${\cal \widehat{L}}$ and ${\cal
\widehat{Q}}$ [16,17]. Selecting an allowable system of
homogeneous coordinates $\breve{z}_\alpha$ ($\alpha$=1,2,3,4) in
such a way that
\begin{equation}
{\cal \widehat{L}}:~~ \breve{z}_1 = 0 = \breve{z}_2,
\end{equation}
\begin{equation}
{\cal \widehat{Q}}:~~ \breve{z}_4 = 0 = - 2\breve{z}_1\breve{z}_2
+ \breve{z}_1\breve{z}_3 + \breve{z}_2\breve{z}_3 \equiv {\cal
\overline{C}},
\end{equation}
\begin{equation}
\widehat{B}_1:~~ \varrho\breve{z}_\alpha = (0,a,b,c),~~a,c \neq 0,
\end{equation}
\begin{equation}
\widehat{B}_2:~~ \varrho\breve{z}_\alpha = (f,0,g,h),~~f,h \neq 0,
\end{equation}
\begin{equation}
\widehat{B}_3:~~ \varrho\breve{z}_\alpha = (k,k,l,m),~~k,l,m \neq
0,
\end{equation}
\begin{equation}
D_1:~~ \varrho\breve{z}_\alpha = (0,0,1,0),
\end{equation}
\begin{equation}
D_2:~~ \varrho\breve{z}_\alpha = (0,0,0,1),
\end{equation}
where $\varrho$ is a non-zero proportionality factor, and
assuming, without any substantial loss of generality, that
\begin{equation}
\frac{a}{c}=\frac{f}{h}=2\frac{k(l-k)}{lm} \equiv - \Theta,
\end{equation}
the web in question is given by
\begin{equation}
{\cal W}(\eta_\alpha):~ \eta_1 \breve{z}_1 \breve{z}_4 \left(
\frac{k(g\breve{z}_1 - f\breve{z}_3)}{fl-gk}+ \breve{z}_2 \right)+
\eta_2 \breve{z}_2 \breve{z}_4 \left(\breve{z}_1 +
\frac{k(b\breve{z}_2 - a\breve{z}_3)}{al-bk} \right)
 + \eta_3 \breve{z}_1 \overline{\cal D}+ \eta_4 \breve{z}_2 \overline{\cal D} = 0,
\end{equation}
with
\begin{equation}
\overline{\cal D} \equiv -2\breve{z}_1\breve{z}_2 +
\breve{z}_1\breve{z}_3 + \breve{z}_2\breve{z}_3 + \Theta
\breve{z}_3\breve{z}_4 = {\cal \overline{C}}+ \Theta
\breve{z}_3\breve{z}_4,
\end{equation}
and $\eta_\alpha \in \Re$. This web generates the following
Cremona transformation [16]
\begin{equation}
\varrho \breve{z}'_1 = \breve{z}_1 \breve{z}_4 \left(
\frac{k(g\breve{z}_1 - f\breve{z}_3)}{fl-gk}+ \breve{z}_2 \right),
\end{equation}
\begin{equation}
\varrho \breve{z}'_2 = \breve{z}_2 \breve{z}_4 \left(\breve{z}_1 +
\frac{k(b\breve{z}_2 - a\breve{z}_3)}{al-bk} \right),
\end{equation}
\begin{equation}
\varrho \breve{z}'_3 = \breve{z}_1 \overline{\cal D},
\end{equation}
\begin{equation}
\varrho \breve{z}'_4 = \breve{z}_2 \overline{\cal D},
\end{equation}
where $\breve{z}'_\alpha$ are the homogeneous coordinates of an
allowable system in the second (``primed") projective space.

Our forthcoming task is to find the {\it fundamental} elements of
${\cal W}(\eta_\alpha)$. To begin with, we recall [6,16] that the
fundamental element of a Cremona transformation between two
3-dimensional projective spaces is a curve, or a surface, in one
space whose corresponding image in the other space is a single
{\it point};\footnote{Equivalently, a fundamental element
associated with a given homaloidal web of surfaces is, in general,
any curve/surface whose only intersections with a generic member
of the web are the base (i.e., common to all the members) elements
of the latter [16].} the loci of fundamental elements of the same
kind being mapped, in general, into {\it curves}, i.e., {\it
one}-dimensional geometrical objects, of the second space.
Employing Eqs.(11)--(14), it is quite a straightforward task to
spot that in our present case the loci of such elements are the
plane of the conic ${\cal \widehat{Q}}$, the three planes
$\widehat{B}_i{\cal \widehat{L}}$, the quadric $\overline{\cal
D}$=0 and the quadratic cone ${\cal \overline{C}}$=0, for their
images in the second space are indeed curves, namely lines (for
the planes and the quadric) and/or a twisted cubic (for the cone)
[17]. To be more explicit, the plane of ${\cal \widehat{Q}}$ host
a pencil (i.e., linear, single parametrical set) of fundamental
{\it lines} ($\vartheta_{1,2} \in \Re$)
\begin{equation}
{\cal \widetilde{L}}(\vartheta_{1,2}):~~ \vartheta_1 \breve{z}_1 +
\vartheta_2 \breve{z}_2 = 0 = \breve{z}_4,
\end{equation}
whose point of concurrence is the point $D_1$; a line from this
pencil has for its primed counterpart a point of the line
$^{s}{\cal \widehat{L}}':~\breve{z}'_1 = 0 = \breve{z}'_2 $, the
latter being single (hence the superscript ``$s$") on the surfaces
of the inverse homaloidal web. The three planes
$\widehat{B}_i{\cal \widehat{L}}$ contain each a pencil of
fundamental {\it conics} whose four base (i.e., shared by all the
members) points are $D_1$, $D_2$, $\widehat{B}_i$ and $K_i$ -- the
last one being the point, not on ${\cal \widehat{L}}$, in which
the plane in question cuts the conic ${\cal \widehat{Q}}$; in
particular,
\begin{equation}
{\cal \widetilde{Q}}_{i=1}(\vartheta_{1,2}):~~ \breve{z}_1 = 0 =
\vartheta_1 \breve{z}_4(b\breve{z}_2 - a \breve{z}_3) +
\vartheta_2 \breve{z}_3 (\breve{z}_2 + \Theta \breve{z}_4),
\end{equation}
\begin{equation}
{\cal \widetilde{Q}}_{i=2}(\vartheta_{1,2}):~~ \breve{z}_2 = 0 =
\vartheta_1 \breve{z}_4(g \breve{z}_1 - f \breve{z}_3) +
\vartheta_2 \breve{z}_3 (\breve{z}_1 + \Theta \breve{z}_4),
\end{equation}
\begin{equation}
{\cal \widetilde{Q}}_{i=3}(\vartheta_{1,2}):~~ \breve{z}_1 -
\breve{z}_2 = 0 = \vartheta_1 \breve{z}_4(l \breve{z}_2 - k
\breve{z}_3) + \vartheta_2 (-2\breve{z}_{2}^2 +
2\breve{z}_2\breve{z}_3 + \Theta \breve{z}_3\breve{z}_4).
\end{equation}
It is readily verified that a conic of ${\cal
\widetilde{Q}}_{i}(\vartheta_{1,2})$ corresponds to a point of the
line  $^{d}{\cal \widehat{L}}'_i$, where $^{d}{\cal
\widehat{L}}'_{i=1}:~\breve{z}'_1 = 0 = \breve{z}'_3$, $^{d}{\cal
\widehat{L}}'_{i=2}:~\breve{z}'_2 = 0 = \breve{z}'_4$, and
$^{d}{\cal \widehat{L}}'_{i=3}:~\breve{z}'_3 - \breve{z}'_4 = 0 =
(l- gk/f)\breve{z}'_1 - (l-bk/a)\breve{z}'_2$, respectively; all
these lines are double (``$d$") on a generic homaloid of the
inverse web. The intrinsic structure and mutual coupling between
these four pencils are depicted in Figure 1. If one compares this
configuration with the one introduced and studied in detail in
[6], which is associated with a homaloidal web of quadric surfaces
and which reproduces what is macroscopically observed, one finds
that the two configurations are prefect inverses of each other; it
was, among other things, also this feature that motivated us to
examine thoroughly this particular kind of Cremonian universe.

\begin{figure}[t]
\centerline{\includegraphics[width=9.0truecm,clip=]{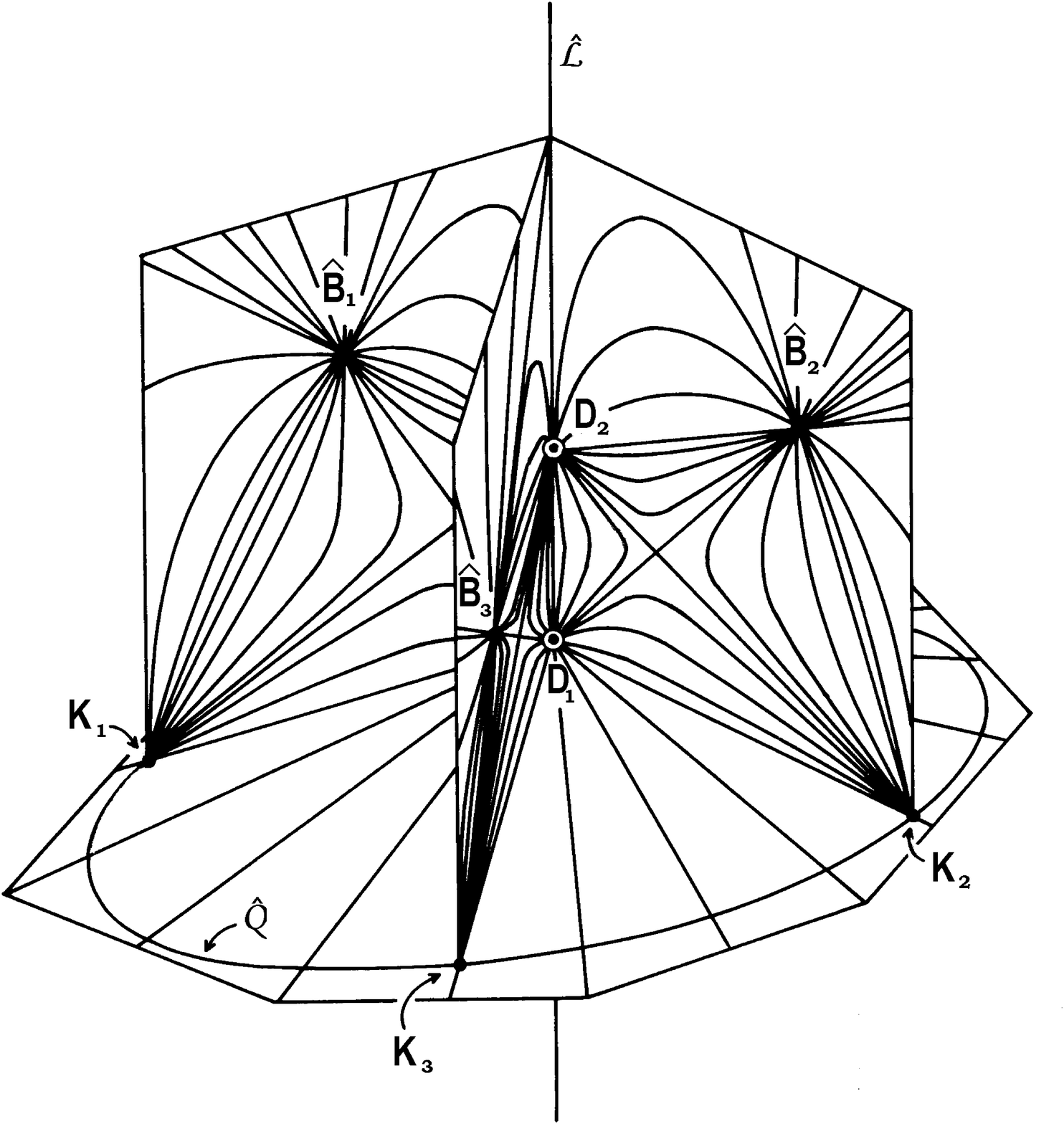}}
\caption{A schematic sketch of the structure of the configuration
of the four ``fundamental" pencils defined by Eqs.\,(15)--(18). In
each pencil, out of an infinite number of its members only several
are drawn. This configuration represents the space-time ``sector"
of the corresponding Cremonian manifold, with three time
dimensions (${\cal \widetilde{Q}}_{i}$) and a single space one
(${\cal \widetilde{L}}$).}
\end{figure}

Now we turn to quadratic loci of fundamental elements. The quadric
$\overline{\cal D}$=0, which is proper (i.e., non-composite) and
ruled (i.e., containing infinity of lines), accommodates a single
parametrical aggregate of fundamental {\it quartics}, i.e., curves
of order four,

\begin{equation}
{\cal  \widetilde{F}}(\vartheta_{1,2}):~ \vartheta_1 \breve{z}_1
\left( \frac{k(g\breve{z}_1 - f\breve{z}_3)}{fl-gk}+ \breve{z}_2
\right)+ \vartheta_2 \breve{z}_2 \left(\breve{z}_1 +
\frac{k(b\breve{z}_2 - a\breve{z}_3)}{al-bk} \right)= 0 =
\overline{\cal D};
\end{equation}
these quartics share the five points $\widehat{B}_i$ ($i$=1,2,3),
$D_1$ and $D_2$, and, in the primed space, they correspond to the
points of the line $^{q}{\cal \widehat{L}}':~\breve{z}'_3 = 0 =
\breve{z}'_4 $, the latter being of multiplicity four (``$q$") on
the inverse homaloids. All the proper quartics in the set are
singular, $D_2$ being their common double point, and, as it is
also obvious from Figure 2, they are genuine space curves. There
are just three composite quartics within this aggregate, each
comprising a pair of conics, namely ($\vartheta \equiv
\vartheta_2/\vartheta_1$)
\begin{equation}
{\cal  \widetilde{F}}(\vartheta=0) \equiv {\cal
\widetilde{F}}_{0}^\odot:~~\breve{z}_1=0= \overline{\cal
D}~~\cup~~ \frac{k(g\breve{z}_1 - f\breve{z}_3)}{fl-gk}+
\breve{z}_2 = 0 = \overline{\cal D},
\end{equation}
\begin{equation}
{\cal  \widetilde{F}}(\vartheta=\infty) \equiv {\cal
\widetilde{F}}_{\infty}^\odot:~~\breve{z}_2=0= \overline{\cal
D}~~\cup~~ \breve{z}_1 + \frac{k(b\breve{z}_2 -
a\breve{z}_3)}{al-bk} = 0 = \overline{\cal D},
\end{equation}
\begin{equation}
{\cal  \widetilde{F}}(\vartheta=\wp) \equiv {\cal
\widetilde{F}}_{\wp}^\odot:~~\breve{z}_1-\breve{z}_2=0=
\overline{\cal D}~~\cup~~ ag\breve{z}_1 + bf\breve{z}_2 -
af\breve{z}_3 = 0 = \overline{\cal D},
\end{equation}
where $\wp \equiv - f(al-bk)/a(fl-gk)$. Figure 2 illustrates the
shape of this aggregate for a generic case where each composite
quartic comprises a pair of {\it proper} conics. (This property
does not hold in our constrained case (see Eq.\,(8)), where one of
the conics of both ${\cal \widetilde{F}}_{0}^\odot$ and ${\cal
\widetilde{F}}_{\infty}^\odot$ is {\it composite} (a line pair).)
\begin{figure}[t]
\centerline{\includegraphics[width=8.0truecm,clip=]{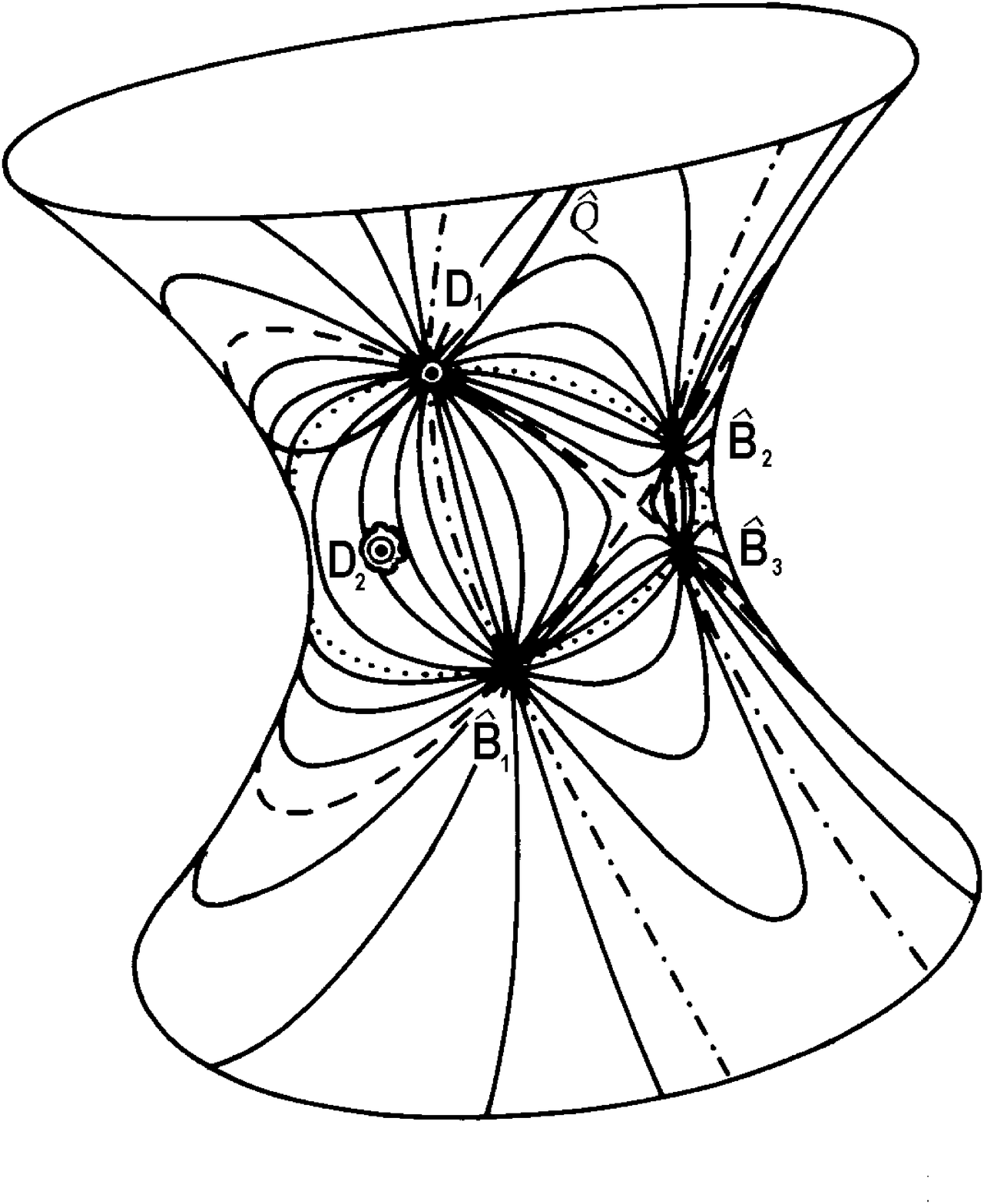}}
\caption{A schematic sketch of the structure of the
singly-infinite set of quartics defined by Eq.\,(19) for  the
generic case where the constraint imposed by Eq.\,(8) is relaxed.
As in the previous figure, only a few quartics are illustrated,
including the composites ${\cal \widetilde{F}}_{0}^\odot$
(dot-dashed), ${\cal \widetilde{F}}_{\infty}^\odot$ (dotted) and
${\cal \widetilde{F}}_{\wp}^\odot$ (dashed).}
\end{figure}

In the case of the quadric cone, ${\cal \overline{C}}$=0, the
fundamental elements are again {\it lines}, forming the following
singly-infinite family
\begin{equation}
{\cal \widetilde{L}}_{{\cal \overline{C}}}(\vartheta):~~
\breve{z}_1 - \vartheta \breve{z}_2 = 0 = (\vartheta+1)\breve{z}_3
- 2\vartheta \breve{z}_2,
\end{equation}
with the parameter $\vartheta$ running through all the real
numbers and infinity as well; for substituting the last equation
into Eqs.\,(11)--(14) yields ($\varsigma \neq 0$)
\begin{equation}
\varsigma \breve{z}'_1= \vartheta \left(\vartheta (\vartheta + 1)
\frac{gk}{fl-gk} + (\vartheta + 1) - 2\vartheta \frac{fk}{fl-gk}
\right),
\end{equation}
\begin{equation}
\varsigma \breve{z}'_2= \vartheta (\vartheta + 1) + (\vartheta +
1) \frac{bk}{al-bk} - 2\vartheta \frac{ak}{al-bk},
\end{equation}
\begin{equation}
\varsigma \breve{z}'_3= 2 \Theta \vartheta^2,
\end{equation}
\begin{equation}
\varsigma \breve{z}'_4= 2 \Theta \vartheta,
\end{equation}
which say that a generic line of ${\cal \widetilde{L}}_{{\cal
\overline{C}}}(\vartheta)$ corresponds indeed to a single point of
the second projective space. The structure of this aggregate can
be discerned from Figure 3.
\begin{figure}[t]
\centerline{\includegraphics[width=8.0truecm,clip=]{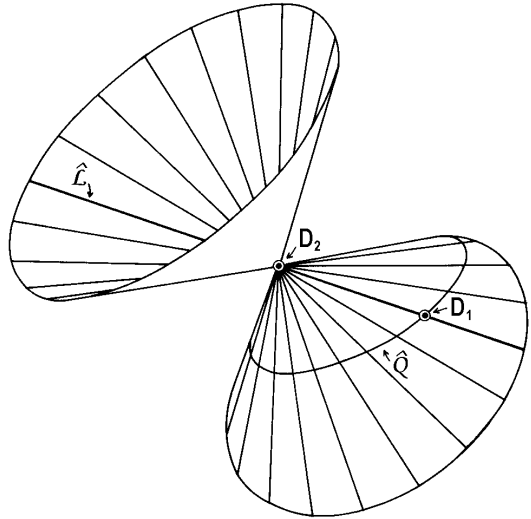}}
\caption{A schematic sketch of the structure of the
singly-infinite set of lines ``sweeping" the quadric cone ${\cal
\overline{C}}$=0. As in the previous two figures, only a finite
number of lines are drawn.}
\end{figure}

At this point we invoke our fundamental ``Cremonian" postulate
[6--10] which says that each single parametrical sets of
fundamental elements generates/represents a unique dimension of a
Cremonian universe, with {\it pencil}-borne aggregates having a
special status of generating space (lines) and time (conics).
Hence, the Cremonian universe under discussion is six-dimensional,
with three dimensions of time (${\cal \widetilde{Q}}_{i}$), one
dimension of space (${\cal \widetilde{L}}$) and two additional
dimensions (${\cal  \widetilde{F}}$ and ${\cal
\widetilde{L}}_{{\cal \overline{C}}}$) of a {\it different}
nature. It is these two extra-dimensions which are of our next
concern.

It is obvious that the ${\cal  \widetilde{F}}$-dimension  bears
more resemblance to time, for its constituents are, like conics,
of non-linear character, whereas the ${\cal \widetilde{L}}_{{\cal
\overline{C}}}$-one, whose elements are lines, is more similar to
the spatial dimension. Yet, there exists a profound difference
between the four pencil-dimensions and these two
``extra-"dimensions. This difference stems from the fact that the
former are all {\it planar} configurations, whereas the latter are
both located on {\it quadratic} surfaces, and acquires its most
pronounced form when a particular Cremonian observer is concerned.
For a Cremonian observer is represented by a line [10,14,15], and
whilst {\it any} line in a three-dimensional projective space is
incident with {\it any} plane (see, e.g., Refs.\,[18,19]), this is
no longer the case for a pair comprising a line and a
non-composite quadric; given any non-composite quadric (whether
proper, or a cone), there exist an infinite number of lines
incident with it, but also an infinity of lines which have no
intersection with this quadric. If we take this incidence relation
as a synonym of the observer's awareness of the particular
dimension, then we see that the four ``planar" dimensions (i.e.,
time and space) will be observed by (accessible to) every
observer, while the two ``quadratic" ones not! In other words, for
each of these two non-planar dimensions, there exist two distinct
groups of the observers; one  comprising observers who perceive
this dimension, the other those who do not. This finding thus
amounts to saying that these two extra-dimensions are observable
only {\it conditionally}.

It is of crucial importance to realize here that the conditional
observability of these extra-dimensions has {\it nothing} to do
with their length (compactification), as no concept of the
measure/metric has so far been introduced into our model. Instead,
it is of a purely algebraic geometrical origin, based solely on
the incidence relations between relevant geometrical objects and
intimately linked with the fact that the ground field of the
background projective space, taken to be that of the real numbers,
is {\it not} algebraically closed; for the {\it geometrical}
problem of finding the common points of a line and a quadric in a
3-dimensional projective space defined over a given ground field
reduces to the {\it algebraic} one of solving/factoring a
quadratic equation in the given field, which is not always
possible unless the field is algebraically closed (see, e.g., Ref.
[20]). This toy-model thus demonstrates that there might exist
extra-dimensions that need not necessarily be
compactified/curled-up to remain unobservable. And this is a truly
serious implication, especially for cosmology and high energy
particle physics.
\\\\
\noindent
{\bf Acknowledgements}\\
I would like to thank Mr. Pavol Bend\'{\i}k for careful drawing of
the figures and Dr. Richard Kom\v z\'{\i}k for the help with
creating their electronic forms. This work was supported in part
by a 2001--2002 NATO/FNRS Advanced Research Fellowship and the
2001--2003 CNR-SAV joint research project ``The Subjective Time
and its Underlying Mathematical Structure." I also wish to
acknowledge the support received from a 2004 ``S\'ejour
Scientifique de Haut Niveau" Physics Fellowship of the French
Ministry of Youth, National Education and Research (No.
411867G/P392152B).

\end{document}